\documentclass[sigconf, nonacm]{acmart}

\usepackage[dvipsnames]{xcolor}

\usepackage{amsmath}
\usepackage{amsfonts}
 
\usepackage{amssymb}

\usepackage{algorithmic}
\usepackage{graphicx}
\usepackage{textcomp}
\usepackage{xcolor}
\usepackage{algorithm}
\usepackage{booktabs}
\usepackage[dvipsnames]{xcolor}

\begin{document}
\title{Robust LLM-based Column Type Annotation via Prompt Augmentation with LoRA Tuning}

\author{Hanze Meng}
\affiliation{%
  \institution{University of British Columbia}
  \streetaddress{201-2366 Main Mall}
  \city{Vancouver}
  \state{British Columbia, Canada}
  \postcode{V6T 1Z4}
}
\email{hmeng99@cs.ubc.ca}

\author{Jianhao Cao}
\affiliation{%
  \institution{University of British Columbia}
  \streetaddress{201-2366 Main Mall}
  \city{Vancouver}
  \state{British Columbia, Canada}
  \postcode{V6T 1Z4}
}
\email{jhcao@cs.ubc.ca}

\author{Rachel Pottinger}
\affiliation{%
  \institution{University of British Columbia}
  \streetaddress{201-2366 Main Mall}
  \city{Vancouver}
  \state{British Columbia, Canada}
  \postcode{V6T 1Z4}
}
\email{rap@cs.ubc.ca}

\begin{abstract}
Column Type Annotation (CTA) is a fundamental step towards enabling schema alignment and semantic understanding of tabular data. Existing encoder-only language models achieve high accuracy when fine-tuned on labeled columns, but their applicability is limited to in-domain settings, as distribution shifts in tables or label spaces require costly re-training from scratch. Recent work has explored prompting generative large language models (LLMs) by framing CTA as a multiple-choice task, but these approaches face two key challenges: (1) model performance is highly sensitive to subtle changes in prompt wording and structure, and (2) annotation F1 scores remain modest. 
A natural extension is to fine-tune large language models. However, fully fine-tuning these models incurs prohibitive computational costs due to their scale, and the sensitivity to prompts is not eliminated. In this paper, we present a parameter-efficient framework for CTA that trains models over prompt-augmented data via Low-Rank Adaptation (LoRA)\footnote{Our code is available at \url{https://github.com/fripSideMeng/PACTA}.}. Our approach mitigates sensitivity to prompt variations while drastically reducing the number of necessary trainable parameters, achieving robust performance across datasets and templates. Experimental results on recent benchmarks demonstrate that models fine-tuned with our prompt augmentation strategy maintain stable performance across diverse prompt patterns during inference and yield higher weighted F1 scores than those fine-tuned on a single prompt template. 
These results highlight the effectiveness of parameter-efficient training and augmentation strategies in developing practical and adaptable CTA systems.
\end{abstract}


\maketitle

\section{Introduction}
\label{sec:intro}

Despite the abundance of tabular data on the web, most tables are published with limited metadata, leaving column semantics implicit and often ambiguous. Therefore, augmenting the metadata available about columns is a fundamental task for making tables interpretable and useful in downstream applications such as schema matching~\cite{DBLP:journals/vldb/RahmB01}, knowledge base population, and data integration~\cite{10.1145/543613.543644}. 
Column type annotation (CTA), which annotates columns with additional information, is a critical prerequisite for enabling automated semantic alignment and interpretation of tabular data.

\begin{figure}[!htbp]
    \centering
    \includegraphics[width=0.9\linewidth]{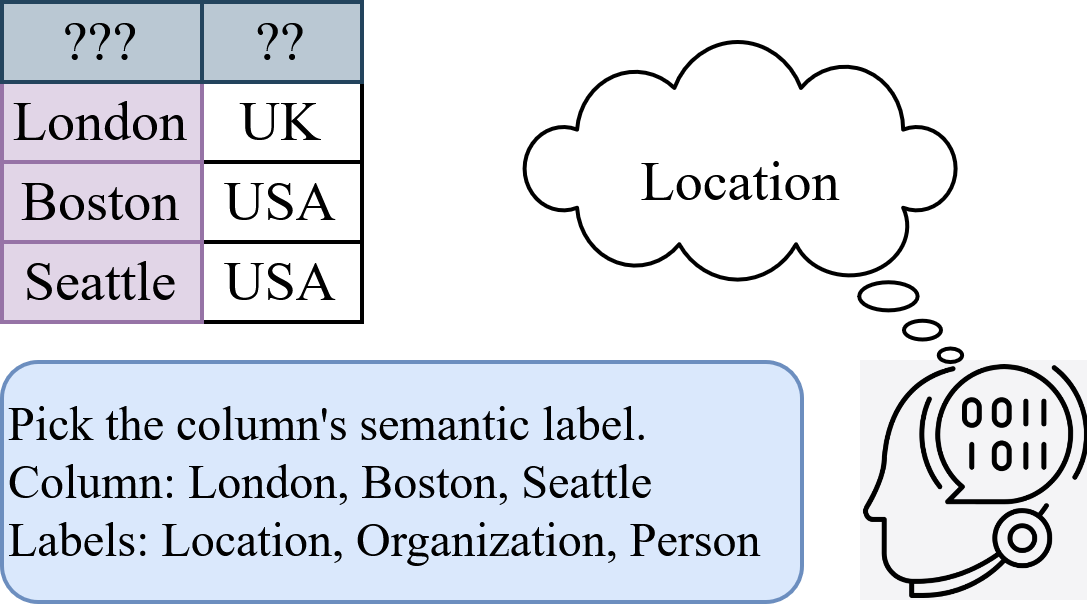}
    \caption{An example of the prompt-based Column Type Annotation (CTA) task. 
    Given a table with unlabeled columns, column entries are presented along with 
    candidate semantic labels in a prompt to a language model. 
    The model predicts the proper semantic type, which in this case is ``Location''.}
    \label{fig:running-example}
\end{figure}

Figure~\ref{fig:running-example}
gives an illustrative case of column type annotation. The input table contains city names such as London, Boston, and Seattle, alongside country codes. The task is to annotate the unlabeled column marked in purple by assigning it the most appropriate semantic type from a candidate set of labels, such as \emph{Location}, \emph{Organization}, or \emph{Person}. Importantly, these semantic types go beyond primitive data types (e.g., \texttt{string}, \texttt{integer}, or \texttt{float}) and instead aim to capture the higher-level meaning or real-world category represented by the column values.

Manual annotation of column semantic types is labor-intensive, error-prone, and unscalable, especially when the number of tables and domains grows. Human annotators must rely on implicit cues such as entry patterns, header keywords, or surrounding context; thus, results are inconsistent from person to person. In large-scale data lakes or web-extracted tables, the absence of standardized metadata means that annotating even a modest fraction of tables requires substantial human effort. Automating this process is essential not only for efficiency but also for enabling downstream applications that rely on consistent and accurate column semantics at scale. Early work used neural networks that leveraged statistical features to classify semantic types~\cite{10.1145/3292500.3330993, 10.14778/3407790.3407793}. The advent of transformer-based architectures has enabled researchers to leverage pre-trained encoder-only language models, leading to strong performance on annotation tasks~\cite{10.1145/3542700.3542709, 10.1145/3514221.3517906}. 

More recently, prompting generative large language models (LLMs) has emerged as an alternative to previous methods. By casting CTA as an instruction-following task, these models can, in principle, adapt to new domains without explicit re-training. Yet, in practice, such prompt-based methods remain limited in performance. Even with tailored pre- and post-processing techniques proposed in~\cite{10.14778/3665844.3665857}, models cannot achieve accuracy greater than 66\% over the SOTAB dataset~\cite{Korini2022SOTABTW}, a challenging benchmark composed of real-world web tables.
More critically, current LLMs are highly sensitive to prompts: minor changes in wording or instruction formatting can cause significant fluctuations in performance \cite{maus2023blackboxadversarialprompting, zhuo-etal-2024-prosa}. To illustrate the sensitivity of models to prompt phrasing, consider the following two prompt designs when annotating the column:

\begin{enumerate}
    \item \texttt{INSTRUCTION: Select the option which best describes the input. INPUT: [`on-schedule'] OPTION: ..., Status Type, ..., Date, ... Answer:}
    \item \texttt{Pick the column's label. Column: [`on-schedule']. Labels: ..., Status Type, ..., Date, ... Output:}
\end{enumerate}
While each formulation yields deterministic outputs under fixed generation settings, the current state-of-the-art prompt-based CTA framework ArcheType~\cite{10.14778/3665844.3665857} produces inconsistent outputs across these two semantically equivalent instructions,
 answering the first prompt with ``Status Type'' and the second with ``Date''. This discrepancy underscores the model's sensitivity to prompt formulation, posing a serious obstacle to reliable deployment: such models fail to generalize across prompt designs, and thus there is no universally optimal prompt that performs well across all datasets and models, implying that identifying an effective formulation often depends on ad-hoc trial and error. Consequently, achieving stability across semantically equivalent prompts is not a secondary objective but a prerequisite for CTA systems. While one might hope that full-model supervised fine-tuning~\cite{ouyang2022training} could alleviate this sensitivity, our experiments show that the problem persists and that such fine-tuning incurs substantially higher computational costs than encoder-only baselines. These factors together render such approaches impractical for real-world CTA deployment.

This paper tackles the challenges of ensuring consistent and accurate responses to semantically equivalent prompts, which currently produce widely varying outputs in state-of-the-art systems.
We propose a prompt augmentation strategy that systematically constructs multiple prompts from each column to fine-tune a large language model, exposing the model to variant but semantically equivalent prompt inputs during training. Additionally, to avoid incurring the high computational costs of full model fine-tuning, we build a parameter-efficient framework for CTA that leverages \emph{Low-Rank Adaptation (LoRA)}~\cite{hu2022lora} to fine-tune models using only a small fraction of their parameters. By combining LoRA tuning with prompt augmentation, our approach stabilizes performance across prompt variants and lowers the training overhead, thus enabling improved robustness in a form that is practical for real-world deployment.

At a high level, our contributions are threefold:
\begin{itemize}
    \item We identify common structural patterns across CTA prompts and propose a training dataset augmentation method that introduces prompt variations by sampling cell values from each column instance and applying multiple template patterns. This augmentation strategy helps to reduce prompt sensitivity and improve generalization during fine-tuning.
    \item We conduct experiments on two real-world web table benchmarks, VizNet~\cite{10.1145/3290605.3300892} and SOTAB~\cite{Korini2022SOTABTW}, demonstrating that our prompt augmentation method consistently enables models to achieve stable performance across different prompt templates and improves weighted F1 scores over single-prompt-template fine-tuning. Moreover, fine-tuning on only one-third of the original training data with our augmentation strategy yields performance comparable to full-data single-prompt-template fine-tuning, highlighting the data efficiency benefits introduced by prompt augmentation.
    \item We show that LoRA fine-tuning is sufficient to significantly improve the performance of prompt-based CTA while requiring far fewer trainable parameters and, consequently, less training time.
\end{itemize}
To our knowledge, this is the first work to integrate prompt-oriented augmentation with LoRA fine-tuning~\cite{hu2022lora} for addressing prompt sensitivity and enhancing label prediction performance in CTA tasks.
The rest of this paper is organized as follows.
Section~\ref{sec:preliminaries} introduces the necessary preliminaries, including the definitions of prompt-based CTA, LoRA fine-tuning, and prompt sensitivity.
Section~\ref{sec:related} provides an overview of the related work.
Section~\ref{sec:loraft} describes the overall fine-tuning pipeline, analyzes common patterns across different prompt templates, and presents our augmentation strategy. 
Section~\ref{sec:experiments} reports experimental results and analysis on real-world web table benchmarks. Finally, Section~\ref{sec:conclusion} concludes with a discussion of implications and future directions.
\section{Preliminaries}
\label{sec:preliminaries}

In this section, we formalize the key concepts underlying our framework: 
prompt-based column type annotation (CTA)~\cite{column_type_annotation_using_chatGPT}, prompt sensitivity~\cite{sclar2024quantifying}, and low-rank adaptation (LoRA)~\cite{hu2022lora}.
\subsection{Prompt-based Column Type Annotation}
\label{subsec:cta}

\subsubsection{Column Type Annotation (CTA)}
The goal of column type annotation is to assign a semantic type to each unlabeled column in a table, selected from a predefined set of labels provided by the user or sourced from a knowledge graph~\cite{10.1007/978-3-030-49461-2_30}. Requiring this label set as input is reasonable, since many practical scenarios, such as enterprise data systems or curated benchmarks, operate within a known set of semantic types tailored to specific tasks.
Formally, we assume a set of relational tables, where a table $T$ is represented as a tuple of columns $(c_1, \ldots, c_n)$. Each column $c_i$ is an ordered list of cell values $v_i$, where each $v_i$ is textual or numeric content. We further assume a predefined finite set of semantic labels $\mathcal{L}$, which constitutes the target label space. The task of CTA is to learn a mapping \( f: c_i \mapsto \ell \) that assigns each unlabeled column \( c_i \) a semantic label \( \ell \) from $\mathcal{L}$. 

Take the table in Figure~\ref{fig:running-example} as an example: given the three labels available, a column containing values such as ``London,'' ``Boston,'' and ``Seattle'' should be annotated as \emph{Location}.
This task is essential for downstream applications such as schema matching, entity linking, and table integration.

\subsubsection{Prompts for Large Language Models}
Large language models (LLMs) are neural networks pre-trained on massive corpora to learn general-purpose linguistic and semantic representations. Most modern LLMs are based on the transformer architecture~\cite{vaswani2017attention} and trained using an autoregressive objective: they learn to generate text by predicting each token conditioned on all previous tokens in the input sequence, where tokens are subword units from the model's vocabulary. 

During \emph{inference}, which is the process of using the model to make predictions without updating its weights, the model generates its output sequentially, predicting each token based on the input prompt and all previously generated tokens. Importantly, the model parameters remain frozen throughout this process. This enables a lightweight form of adaptation known as \emph{prompting}, where task-specific input is expressed in natural language and the model is queried to produce a textual output.
For example, a prompt such as ``What is the semantic type of the column containing values of `\textit{Canada}', `\textit{Brazil}', and `\textit{Japan}'?'' can be used to elicit a model-generated label such as ``\textit{country}''.
This allows pre-trained LLMs to be repurposed across a wide range of tasks, with no additional fine-tuning required. Prompting thus serves as an interface between general pre-training and specific task adaptation. 

In general, a prompt can be thought of as composed of three necessary components: 
(i) a \emph{task description} specifying what the model is expected to do, 
(ii) an \emph{input context} that may provide the data instance(s) to operate on, and 
(iii) an \emph{output specification} constraining the form or space of possible answers. As in Figure~\ref{fig:running-example}, the task description is given by the instruction ``Pick the column's semantic label.'' The input context consists of the sampled column values [``London'', ``Boston'', ``Seattle''], and the output specification is the label space [``Location'', ``Organization'', ``Person''] from which the model must choose.
Conditioned on the prompt, the model generates a response as the task output. Prompts therefore function as the interface between general-purpose pre-training and the requirements of a specific task.

\subsubsection{Prompt-based CTA}
\label{subsubsec:cta_prompt_comp}
In the context of prompting, the CTA task is reformulated as a text-to-text generation problem: the model is prompted with natural language describing the input (i.e., column values and label options), and is expected to generate the correct semantic label as output~\cite{column_type_annotation_using_chatGPT}. Prompts in CTA are constructed by combining three core components:
\begin{itemize}
    \item a \emph{task instruction} $t_i$ that tells the model what to do.
    \item a \emph{sampled column} $s_c$ containing representative entries from the target column.
    \item a \emph{label space} $s_\ell$ listing all candidate semantic types.
\end{itemize}
The model receives the textual encoding of $(t_i, s_c, s_\ell)$ and generates an output string $\hat{\ell} \in \mathcal{L}$, typically matching one of the given labels. Different prompt templates vary in how these components are phrased, formatted, or ordered.

\subsection{Prompt Sensitivity}
\label{subsec:prompt_sensitivity}
Prompt sensitivity refers to the phenomenon where small variations in prompt design lead to large and often unpredictable shifts in model output \cite{maus2023blackboxadversarialprompting,zhuo-etal-2024-prosa}. In the context of CTA, such variations may include rewording the instruction (e.g., ``Classify the column'' vs. ``Pick the column's label''), changing the order in which the label set and sampled column appear, or injecting auxiliary tokens such as ``INSTRUCTION:'' to signal the start of specific prompt components. These changes, while semantically equivalent, can result in remarkable performance differences even when evaluated on the same data using the same model.

This sensitivity issue makes model deployment brittle across datasets or user environments. Importantly, because the optimal prompt is often dataset- and model-specific, users must rely on costly trial and error to identify which template works best—an impractical process when generalizing across many tables or switching models. Even fine-tuned LLMs do not eliminate this instability, as we show later in Section~\ref{sec:experiments}.

\subsection{Low-Rank Adaptation (LoRA)}
\label{subsec:lora}

\subsubsection{Beyond prompting} 
While prompting offers a lightweight mechanism for adapting LLMs to new tasks, prior work~\cite{10.14778/3665844.3665857} has shown that inference-only approaches yield suboptimal performance on CTA in spite of customized pre- and post-processing algorithms, indicating the need for additional task-specific adaptation beyond pre-training.
Therefore, a natural alternative is to update the model parameters through fine-tuning, which can lead to stronger task-specific performance. However, full fine-tuning requires updating all parameters of the model, which is prohibitively costly for modern LLMs with billions of parameters.

\subsubsection{Parameter-efficient tuning} 
To reduce the cost of adapting large pre-trained language models to specific tasks, parameter-efficient fine-tuning methods have been proposed as an alternative to full-model fine-tuning. Instead of updating all of the model's parameters, which can be computationally expensive and memory-intensive, these methods keep the pre-trained model weights frozen and introduce a small number of task-specific trainable parameters. This greatly reduces the training time and memory footprint, while still enabling effective task adaptation. 

Among these techniques, \emph{Low-Rank Adaptation (LoRA)}~\cite{hu2022lora} has emerged as a particularly popular and effective approach, especially for tuning large language models under resource constraints.

\begin{figure}[!htbp]
    \centering
    \includegraphics[width=0.8\linewidth]{figures/lora.png}
    \caption{Illustration of Low-Rank Adaptation (LoRA)~\cite{hu2022lora}. 
    Instead of updating the full matrix $W$, LoRA freezes $W$ and learns two smaller matrices $A$ and $B$ with rank $r \ll d$. 
    This reduces the number of trainable parameters by orders of magnitude while retaining the pre-trained knowledge in $W$.}
    \label{fig:lora}
\end{figure}

\subsubsection{Low-Rank Adaptation} 
Figure~\ref{fig:lora} illustrates the high-level idea of LoRA. Given a large matrix of pretrained weights $W \in \mathbb{R}^{d \times d}$, directly updating it during fine-tuning can be computationally expensive. LoRA addresses this computational cost issue by keeping $W$ frozen and introducing a low-rank adaptation for efficient weight updates through two smaller matrices: $A \in \mathbb{R}^{d \times r}$ and $B \in \mathbb{R}^{r \times d}$, where $r \ll d$. The weight update is then computed as $\Delta W = A \times B$, and the adapted weights are obtained as $W + \Delta W$ at a much lower computational cost. In other words, instead of learning a full \( d \times d \) weight update matrix, LoRA learns a pair of much smaller matrices whose product approximates the desired update.

During fine-tuning, only \( A \) and \( B \) are optimized, while the original weight matrix \( W \) remains frozen. This significantly reduces the number of trainable parameters from \( d^2 \) to \( 2rd \), and allows models to be adapted efficiently to new tasks with minimal additional computation or storage.

\subsubsection{Why LoRA matters for Prompt-based CTA} 
Robustness to prompt and domain variation is essential for CTA, yet the cost of fully fine-tuning large models for every new combination of dataset and prompt makes such an approach impractical. LoRA~\cite{hu2022lora} offers an efficient alternative: it enables large models to be adapted efficiently with only a fraction of the training cost, while preserving the flexibility of swapping in or out the lightweight adapted parameters. To the best of our knowledge, despite its growing popularity in other NLP domains, LoRA has not been systematically explored for CTA tasks.
In this work, we show that LoRA provides an effective and scalable solution for adapting large generative models to CTA, offering strong performance with reduced computational overhead.

\section{Related Work}
\label{sec:related}

\begin{figure*}[t]
    \centering
    \includegraphics[width=0.9\textwidth]{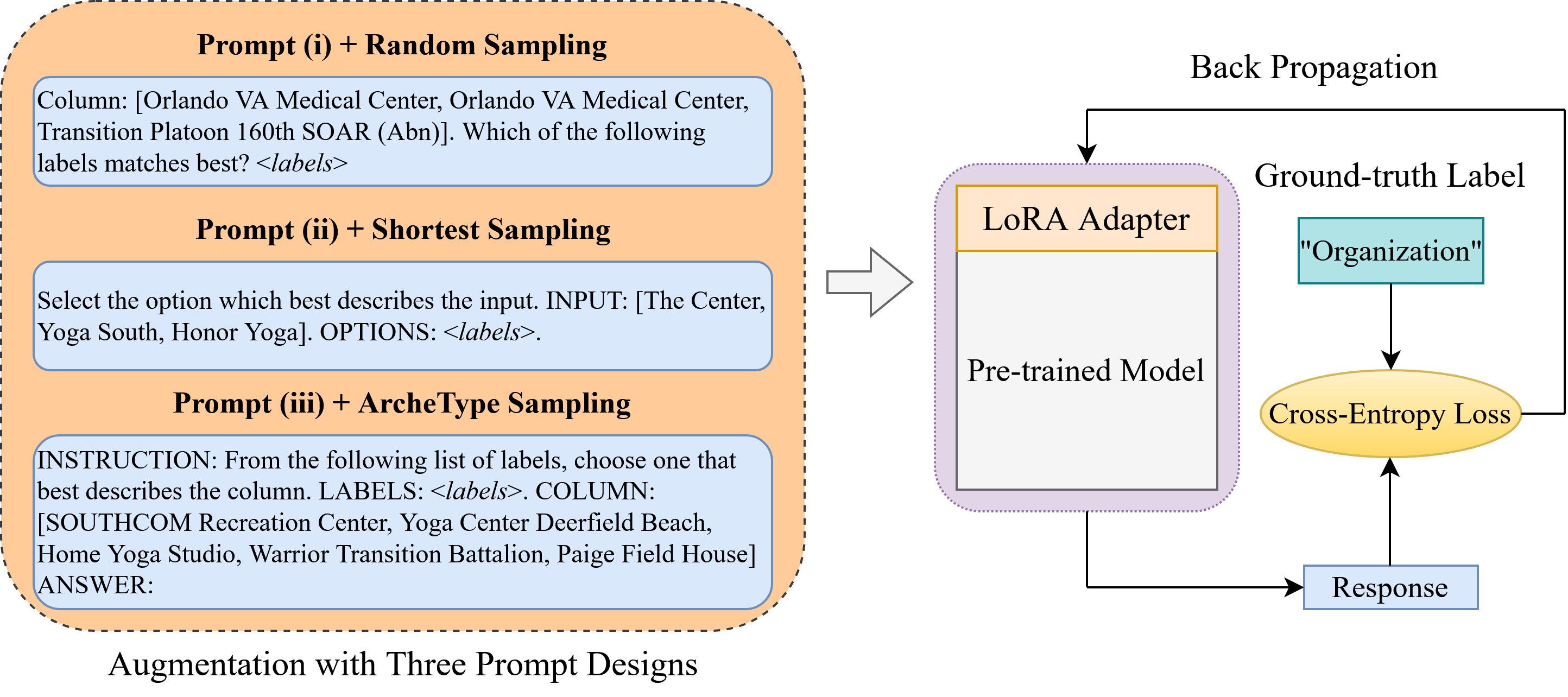}
    \caption{
        Overview of our parameter-efficient fine-tuning pipeline for prompt-based column type annotation (CTA). 
        Given a column, we instantiate the three template patterns to form concrete prompts with different column entry sampling strategies: \textbf{Random Sampling}, \textbf{Shortest Sampling}, and \textbf{Archetype Sampling}, 
        each encoding a different contextual view. 
        These prompts are passed into a frozen pre-trained language model augmented with a trainable \textbf{LoRA Adapter}. 
        The model's prediction (\texttt{Response}) is compared with the ground-truth label (``\texttt{Organization}'') via a cross-entropy loss, whose gradients are backpropagated to update the LoRA adapter. 
        Only the weights in the LoRA adapter are updated during fine-tuning, leaving the base model untouched. 
        This design enables efficient adaptation and prompt-based augmentation without modifying the underlying model. 
    }
    \label{fig:lora-pipeline}
\end{figure*}

\subsection{Prompt-based CTA Methods}
\label{subsec:prompt_based_cta}
An emerging line of research approaches column type annotation (CTA) by prompting pre-trained large language models with instructions, offering a flexible alternative to supervised training. TabLLM \cite{tabllm} explores how to represent tabular data within prompts for classification-based tasks. Extending this line of research on CTA, Korini and Bizer~\cite{column_type_annotation_using_chatGPT} systematically examined prompt design variations in closed-source models, analyzing the effects of instructional phrasing, conversational roles, and few-shot in-context learning~\cite{dong2024surveyincontextlearn}. In a subsequent study, Korini and Bizer~\cite{10.1007/978-3-032-05281-0_8} further supplemented the prompts with LLM-generated column-type definitions as contextual information. RACOON \cite{racoon} improves LLM-based column type annotation by enriching prompts with contextual signals derived from a knowledge graph. Given that generative LLMs can produce inaccurate or irrelevant outputs which are commonly referred to as hallucinations \cite{DBLP:journals/corr/abs-2309-01219}, post-processing is often required. To mitigate this, Chorus \cite{10.14778/3659437.3659461} proposes an anchoring technique that enforces constraints during decoding and substitutes invalid outputs with semantically similar labels.

Table-GPT \cite{10.1145/3654979} proposed an instruction-tuning approach by fine-tuning LLMs with instructions encompassing multiple table understanding objectives. This fine-tuning paradigm enables the model to internalize both tabular semantics and task-specific instructions, allowing it to move beyond merely responding to prompts and to achieve improved performance on specific tasks. ArcheType \cite{10.14778/3665844.3665857} is another CTA framework that supports fine-tuning and introduces a structured LLM-based pipeline comprising context sampling, prompt construction, model querying, and label correction. While fine-tuning large models is typically resource-intensive, we explore a LoRA-based fine-tuning strategy that enables parameter-efficient adaptation, making the approach both practical and scalable for models to internalize task-specific signals and generalize more consistently across variations in input phrasing.

\subsection{Prompt Sensitivity in LLMs}
\label{subsec:prompt_sensitivity_llm}
Large language models are highly sensitive to prompt formatting and option ordering~\cite{maus2023blackboxadversarialprompting, zhuo-etal-2024-prosa}.
Even minor variations such as rewording instructions, changing punctuation, or reordering answer options and few-shot examples can lead to significant shifts in model predictions. Maus et al.~\cite{maus2023blackboxadversarialprompting} leveraged this behavior to generate adversarial prompts that deceive LLMs. Zhao et al.~\cite{zhao2021calibrate} documented this brittleness of prompt sensitivity when prompting on the GPT-3 model~\cite{10.5555/3495724.3495883} for text classification, showing that few-shot learning performance varies dramatically depending on example selection and prompt format. Pezeshkpour and Hruschka~\cite{pezeshkpour-hruschka-2024-large} demonstrated that LLMs are sensitive to the order of options in multiple-choice questions presented in the prompt. Mizrahi et al.~\cite{mizrahi-etal-2024-state} found that different instruction templates can lead to substantially different performance in LLMs. Similarly, Kojima et al.~\cite{10.5555/3600270.3601883} observed that inserting trigger phrases such as \textit{``Let's think step by step''} can elicit multi-step reasoning behavior and significantly improve performance on reasoning tasks, highlighting the importance of phrasing. \textsc{FormartSpread}~\cite{sclar2024quantifying} further investigated prompt sensitivity by evaluating semantic-meaning-preserving prompt variants. They found that open-source LLMs like Llama~\cite{touvron2023llama} exhibit accuracy swings of up to 76 percentage points solely due to formatting differences. Zhuo et al.~\cite{zhuo-etal-2024-prosa} investigated prompt sensitivity at the instance level, where an LLM may perform well on one instance with a given prompt but poorly with another, while exhibiting the opposite behavior on a different instance.

Prompt sensitivity remains a fundamental challenge for the reliable deployment of generative LLMs. Prior work in the natural language processing community has explored several strategies to mitigate this issue. One line of research fine-tunes models on multiple instruction formulations to alleviate sensitivity~\cite{chung2024scaling}. Other approaches include applying loss regularization across different prompt variants to encourage consistent predictions~\cite{yan-etal-2024-contrastive, qiang-etal-2024-prompt} and sampling model responses over perturbed prompts~\cite{cox2025mapping} in conventional NLP tasks. Following the introduction of LLM-based approaches for the CTA task~\cite{column_type_annotation_using_chatGPT, 10.14778/3665844.3665857, 10.14778/3659437.3659461}, we present the first systematic investigation of prompt sensitivity in CTA. In this work, we demonstrate that prompt-augmentation strategies, although often overlooked in this setting, can substantially improve model stability across prompt variants.
\section{Fine-tuning Paradigm}
\label{sec:loraft}

In this section, we present the overall architecture of our parameter-efficient fine-tuning pipeline for prompt-based Column Type Annotation (CTA). To the best of our knowledge, this is the first work to leverage \textbf{prompt-oriented augmentation} in conjunction with \textbf{LoRA} fine-tuning~\cite{hu2022lora} to address the challenges of prompt sensitivity and label prediction performance in prompt-based CTA tasks.
We begin in Section~\ref{subsec:training} by describing the full training pipeline, where LoRA adapters are trained using outputs from diverse prompt formulations while keeping the backbone model frozen. Section~\ref{subsec:decomposition} details how we decompose prompt templates into three components and introduce variations in prompt patterns to construct prompt variants as additional training examples based on the same column. Finally, Section~\ref{subsec:augmentation} demonstrates how prompt decomposition yields a prompt-based augmentation strategy that exposes the model to varying contextual views of the same column, thereby making it more robust and generalizable.

\subsection{Training Pipeline}
\label{subsec:training}
Our framework incorporates prompt formatting to augment training instances and performs fine-tuning of the base model using LoRA adapters, as illustrated in Figure~\ref{fig:lora-pipeline}.
We define each training instance as a prompt instantiated from a template by filling in column values and the corresponding semantic label. The prompt serves as the model input, and the label provides the expected output during training.
We adopt a decoder-based language model as the backbone. As far as we are aware, this is the first framework that augments fine-tuning prompts grounded in tabular data and explicitly addresses prompt sensitivity in CTA, while maintaining computational efficiency through parameter-efficient fine-tuning.

\paragraph{Prompt Formatting} Each column is paired with multiple prompt templates that differ in instruction wording, enabling the model to generate consistent responses for label prediction while remaining robust to prompt variation. Each prompt template is materialized into a fine-tuning instance by filling in different values sampled from the same column, and then fed to the model to generate the predicted label string. In prompt-based CTA, we treat the task as a generation problem and minimize the loss between the model's response to a prompt and the target text sequence (i.e., the ground truth label).

\paragraph{LoRA Adaptation} To fine-tune the model efficiently, we insert lightweight LoRA modules into specific parts of its internal layers so that we modify only a small number of parameters while keeping the original pre-trained weights unchanged. The additional weights can be merged into the model or removed entirely at inference time, offering flexibility for deployment.

\paragraph{Training Objective} Let $\mathcal{D} = \{(x_i, y_i)\}_{i=1}^N$ denote the training dataset. Here, $N$ is the total number of training examples. Each input $x_i$ is a prompt populated with column values and $y_i$ is the corresponding target label, both tokenized into sequences of tokens. The model computes a conditional probability distribution $P(y_i \,|\, x_i; W + \Delta W)$, where $W$ represents the frozen pre-trained weights and $\Delta W$ denotes the LoRA adapter parameters, as illustrated in Figure~\ref{fig:lora}. Although only $\Delta W$ is updated during training, both components contribute to the model's output. The training objective is to minimize the average token-wise cross-entropy loss across the dataset:
\[
\mathcal{J}(\Delta W) = -\frac{1}{N} \sum_{i=1}^{N} \sum_{t=1}^{T_i} \log P(y_i^{(t)} \mid y_i^{(<t)}, x_i; W + \Delta W)
\]
where $T_i$ is the number of tokens in the target label $y_i$, $y_i^{(<t)}$ denotes the first $t-1$ tokens of the target label, and $P(y_i^{(t)} \mid y_i^{(<t)}, x_i; W + \Delta W)$ denotes the model's conditional probability of producing the $t$-th token of label $y_i$, given the input prompt $x_i$ and $y_i^{(<t)}$. Intuitively, this loss encourages the model to generate the correct label one token at a time, using the input prompt and the earlier tokens of the label as context. At each position, the model is trained to predict the next token in the ground-truth label with exact matches, and is penalized whenever its prediction differs.

To make this concrete, consider the following input prompt: \texttt{Pick the column's semantic label. Column: [`on-schedule'] Labels: ..., Status Type, ..., Date, ... Answer:}, and the ground-truth label of the column ``Status Type''. During training, the model is guided to generate this label token by token, using the input prompt as context. At each generation step, it is given the correct beginning of the label so far (``Status'') and learns to continue it with the next correct token (``Type''). This setup prevents small mistakes from snowballing into larger ones, and helps the model build a stable mapping between the given column values and the expected label.
\paragraph{Training Workflow} The process is summarized in Algorithm~\ref{alg:lora-finetune}. For each training batch, we tokenize the input–output pairs, compute the token-level cross-entropy loss, and update the LoRA parameters $\Delta W$ via backpropagation.

\begin{algorithm}[H]
\caption{LoRA Fine-Tuning on augmented prompts}
\label{alg:lora-finetune}
\begin{algorithmic}[1]
\REQUIRE Training set $\mathcal{D} = \{(c_i, \ell_i)\}$, where $c_i$ is a column and $\ell_i$ its semantic label, Set of prompt templates $\mathcal{P}$, Set of all semantic labels $\mathcal{L}$
\STATE Construct augmented prompt set $\tilde{\mathcal{D}} = \{(x, y)\ |\ x=p(c_i, \mathcal{L}),$ $y=\ell_i,\ (c_i, \ell_i) \in \mathcal{D}\}$ by populating a prompt template $p \in \mathcal{P}$ with $c_i$ and $\mathcal{L}$
\FOR{each training step}
    \STATE Sample a batch of $(x, y)$ pairs from augmented set $\tilde{\mathcal{D}}$
    \STATE Tokenize the input prompt $x$ and target label $y$
    \STATE Pass input tokens into the model with LoRA adapter
    \STATE Compute output token probabilities $\hat{y}$
    \STATE Compute the loss $\mathcal{J}(\Delta W)$ between $\hat{y}$ and $y$
    \STATE Back-propagate gradients and update $\Delta W$ only
\ENDFOR

\end{algorithmic}
\end{algorithm}

\subsection{Prompt Template Variants}
\label{subsec:decomposition}

Although the fine-tuning pipeline enables the model to learn from prompt-formatted column data, it remains vulnerable to variations in prompt phrasing, which is the \emph{prompt sensitivity} issue defined in Section~\ref{sec:preliminaries}. This refers to the model's tendency to produce inconsistent predictions when semantically equivalent instructions are presented in different surface forms. Such brittleness can limit the practical reliability of prompt-based CTA systems.

\begin{figure}[t]
    \centering
    \includegraphics[width=0.44\textwidth]{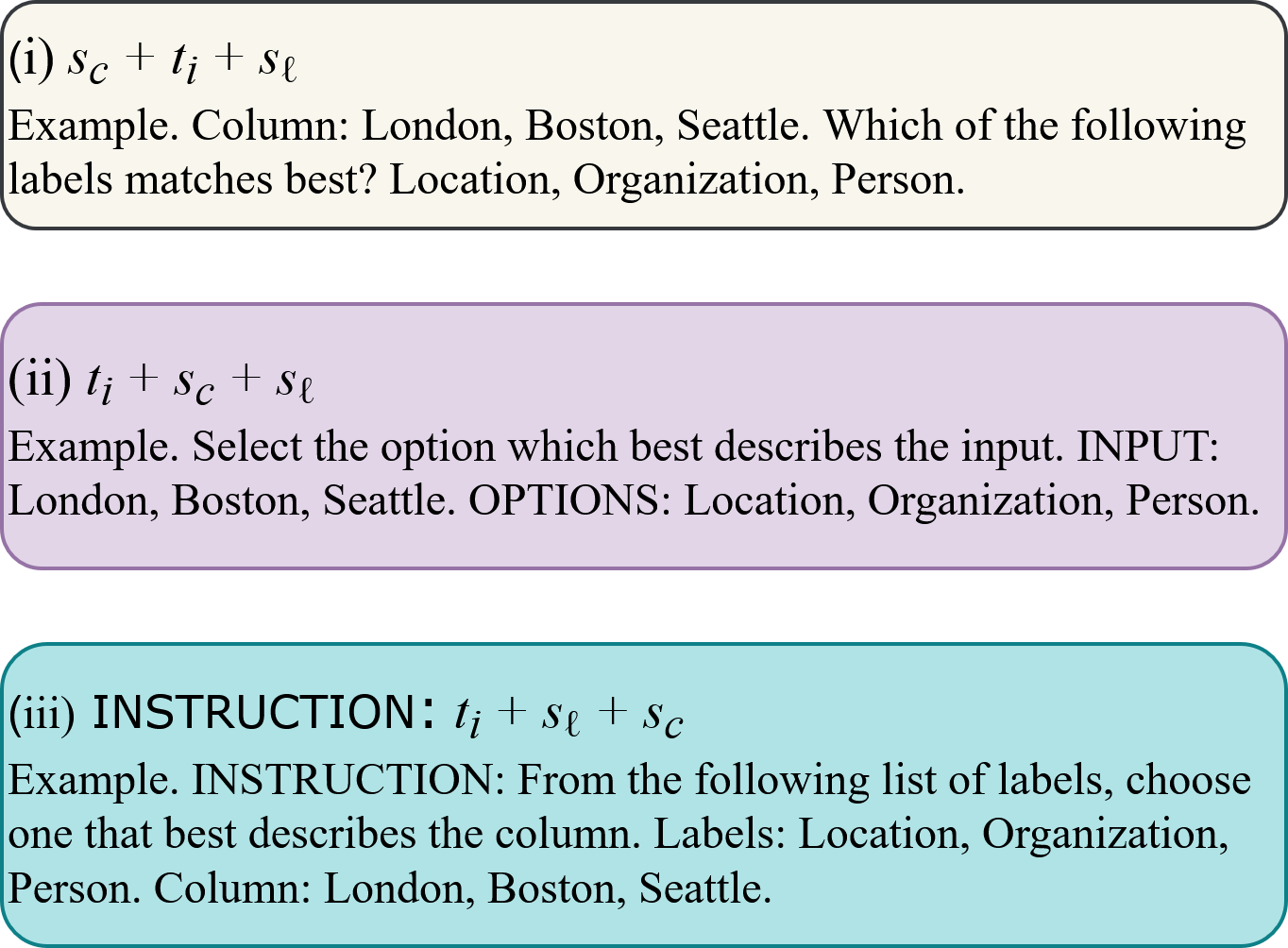}
    \caption{Three prompt patterns identified in prior prompt-based CTA literature. All patterns are populated with the values from the first column in Figure~\ref{fig:running-example}. $t_i$, $s_c$ and $s_\ell$ are the prompt components as defined in Section~\ref{subsubsec:cta_prompt_comp}.}
    \label{fig:aug-strategy}
\end{figure}

Given the almost unbounded space of possible prompt wordings, directly analyzing each surface form is infeasible. Instead, we adopt a pattern-based perspective, observing that practical prompts for CTA conform to a small number of formats. Specifically, we identify three structural patterns that capture the dominant variations in prompt formats used for CTA~\cite{10.14778/3665844.3665857}. Recall from Section~\ref{subsubsec:cta_prompt_comp} that each prompt is composed of three elements: a task instruction ($t_i$), a sampled column ($s_c$), and a label set ($s_\ell$). We observe that existing prompt designs typically follow one of the three prompt patterns listed in Figure~\ref{fig:aug-strategy}.

These forms capture both syntactic permutations and stylistic conventions. By decomposing prompt templates in this way, we obtain a controlled design space for evaluating how models respond to prompt variation. Therefore, our decomposition is not only grounded in empirical patterns observed in prior works, but also enables a more systematic analysis of prompt sensitivity, which is a perspective that, to the best of our knowledge, has not been formally adopted in prior CTA research. This also lays the foundation for our data augmentation strategy, which we discuss in detail in the next subsection.

\subsection{Prompt-Based Data Augmentation with Value Sampling}
\label{subsec:augmentation}

Beyond offering an organized view of variations, prompt template decomposition highlights a key insight: the same column can be rendered using multiple semantically equivalent yet syntactically varied phrasings and patterns. We exploit this insight to develop a \emph{prompt-based data augmentation} strategy for fine-tuning, aimed at improving the model's robustness and generalization. Specifically, this strategy operates by generating multiple views of the same column instance through variations in both prompt patterns and sampled column entries. This not only helps the model fine-tune on variant prompts constructed from the same column, thereby mitigating prompt sensitivity, but also expands the model's exposure to diverse values within each column. Such exposure is especially important when complete inclusion of column entries is infeasible due to the input length limits of generative models. By training on diverse yet semantically consistent prompt formulations, the model sees more informative contents and learns to produce stable predictions across structural variations. Formally, we define the following augmentation strategy:

Let $c$ denote the target column. We first generate subsets of entries from $c$ using the following sampling strategies:

\begin{itemize}
    \item \textbf{ArcheType~\cite{10.14778/3665844.3665857} Sampling}: Favor entries with longer texts, which are likely to contain richer semantic signals.
    \item \textbf{Random Sampling}: Uniformly sample a subset of entries to introduce unbiased diversity.
    \item \textbf{Shortest Sampling}: Select shorter entries to include more examples within the prompt given the token limit.
\end{itemize}

Each sampled subset of entries is embedded into one of the three prompt patterns listed in Figure~\ref{fig:aug-strategy}, yielding multiple augmented training instances for the same column-label pair. Taking the input prompts in Figure~\ref{fig:lora-pipeline} as an example, one prompt follows the \texttt{INSTRUCTION: $t_i$ + $s_\ell$ + $s_c$} template, filled with values from a column selected via ArcheType sampling, while the other two use random samples and the shortest values to materialize the \texttt{$s_c$ + $t_i$ + $s_\ell$} and \texttt{$t_i$ + $s_c$ + $s_\ell$} templates, respectively.

We select these three templates for prompt augmentation to introduce variations in sampled values, task instructions, and the positioning of components within the prompt while preserving the semantic equivalence of the CTA query. These template designs provide a controlled yet representative set of prompt variants with systematic syntactic and structural variations for constructing augmented training instances.

Our prompt-based augmentation strategy leverages phrasing variations in prompt templates and content-level diversity by sampling entry values within each column. By being fine-tuned on multiple variant prompts constructed from the same column, the model becomes more robust to prompt phrasing and better equipped to generalize across varied inputs, producing more consistent responses for column type annotation. Moreover, our method does not require any new labeled columns to create additional training examples for LLMs; instead, it systematically reuses the existing labeled columns to augment training prompts. While motivated by the challenges of CTA, this approach offers a generalizable methodology for enhancing robustness in other prompt-based classification tasks, where model predictions may also be sensitive to superficial input variations. Designing training data to expose such variability can help condition the model toward more stable and task-consistent behavior.
\section{Experiments and Analysis}
\label{sec:experiments}
In this section, we evaluate our framework of fine-tuning pre-trained generative models for column type annotation. 
We begin by describing our setup in Section~\ref{subsec:expr-setup}, including dataset and model selections, fine-tuning settings, evaluation metrics and baseline comparison. Section~\ref{subsec:cta-sensitivity} then examines the performance variability observed in prior prompt-based CTA methods by analyzing how both frozen models and single-prompt-template fine-tuned models respond to different prompt formulations.

We then evaluate our augmentation strategy during fine-tuning in Section~\ref{subsec:aug_improvement}, demonstrating that it enhances model robustness and reduces sensitivity to variations in prompt phrasing and structure, as evidenced by improved weighted F1 scores on unseen prompt variants. 
Next, we compare fine-tuning with a single prompt template with all training data to the augmentation setup that uses only a fraction of the training data, which shows that our augmentation can match the performance of full-data fine-tuning with improved robustness. We also include a case study in Section~\ref{subsec:case_llama} where we update all parameters of the model, in contrast to LoRA-based tuning. The results show that the sensitivity issue persists even with full-parameter fine-tuning when prompt augmentation is not applied, indicating that robustness depends not on the number of updatable parameters but on the model’s exposure to more prompt variants.

\subsection{Experimental Setup}
\label{subsec:expr-setup}
\subsubsection{Baselines}
We compare against ArcheType~\cite{10.14778/3665844.3665857}, a state-of-the-art prompt-based CTA pipeline, 
which leverages generative language models for column type annotation through customized input and output processing, following a single-prompt-template fine-tuning approach. ArcheType presents an inference-oriented framework featuring context sampling, prompt serialization, model querying, and label remapping. We apply the same inference pipeline as ArcheType, by using its official implementation, to predict column labels in our experiments. Although other prompt-based systems have been proposed~\cite{column_type_annotation_using_chatGPT, 10.14778/3659437.3659461}, they rely on closed-source models, making consistent evaluation difficult~\cite{rogers-etal-2023-closed}. We also note that existing works, including ArcheType, do not directly address prompt sensitivity in CTA, which underscores both the necessity and novelty of our investigation.

The parameter-frozen model for direct inference and the single-prompt-template fine-tuning scenario in our experiments align exactly with those in ArcheType, and we use these two settings as baselines to compare against our prompt augmentation approach with LoRA-based fine-tuning in terms of prompt sensitivity.

\subsubsection{Datasets and Models}
We use VizNet~\cite{10.1145/3290605.3300892} and SOTAB~\cite{Korini2022SOTABTW} as our evaluation benchmarks. Both datasets consist of real-world web tables and feature a relatively large number of semantic labels, making them challenging and representative benchmarks for CTA tasks. Specifically, for VizNet, which is a collection of tables originally created to support data visualization, we use the version preprocessed by Sherlock~\cite{10.1145/3292500.3330993}, which is adapted for the CTA task and includes 412,059 columns for training and 137,353 columns each for validation and testing. All three splits share the same set of 78 distinct semantic labels. For SOTAB, the training set consists of 46,790 tables with 130,471 columns, the validation set includes 5,732 tables with 16,840 columns, and the test set comprises 7,026 tables with 15,040 columns. Each split contains the same set of 91 distinct semantic labels.

We select FLAN-T5 Base, FLAN-T5 XXL, and FLAN-UL2~\cite{wei2022finetuned} as the target models for LoRA fine-tuning, and use the Llama 3.1 8B Instruct model~\cite{DBLP:journals/corr/abs-2407-21783} for full-model supervised fine-tuning~\cite{ouyang2022training}. These models are instruction-tuned and decoder-based, making them well-suited for prompt-based CTA tasks. Our model selections also align closely with the choices in ArcheType~\cite{10.14778/3665844.3665857}, ensuring consistency and comparability in evaluation. One notable difference is that while ArcheType uses Llama 2 7B~\cite{touvron2023llama}, we adopt the model from Llama 3.1 8B~\cite{DBLP:journals/corr/abs-2407-21783} due to its more recent pre-training and instruction-tuning advancements, which better reflect the current capabilities of generative models in handling data management tasks. This updated model choice enables us to evaluate whether prompt sensitivity persists even in more modern, instruction-optimized architectures.

\subsubsection{Fine-tuning Settings}

Our experiments are designed to investigate the effectiveness of prompt augmentation in mitigating prompt sensitivity under various fine-tuning settings. Specifically, we examine two factors: (1) the number of training examples used for fine-tuning and (2) the scale of trainable parameters.

To study prompt sensitivity under different data scales, we consider two scenarios: one with limited training examples, where only a small portion of labeled columns from the training set is used to construct prompts for fine-tuning, and another with abundant training examples, where the entire training set is available. To analyze the impact of trainable parameter scale, in addition to the parameter-efficient LoRA fine-tuning approach, we conduct a case study on full-model fine-tuning, where all parameters are updated.

We define the following terminology to distinguish models under different prompt-based fine-tuning settings used throughout the experiments. For clarity, we refer to template (i) in Figure~\ref{fig:aug-strategy} as $p_1$, template (ii) as $p_2$, and template (iii) as $p_3$. 
For inference settings:
\begin{itemize}    
    \item If the model is not fine-tuned and only its pre-trained weights are used, we denote it by the model name followed by the tag ``FROZEN''.
    \item If the model is fine-tuned using LoRA~\cite{hu2022lora}, we denote it by the model name followed by the proportion of training data used and the tag ``LoRA''.
    \item If the model is fully fine-tuned with all parameters updated~\cite{ouyang2022training}, we denote it by the model name followed by the proportion of training data used and the tag ``SFT''.
    \item If only template $p_3$ is used to construct fine-tuning examples, we add a subscript to the fine-tuning method (``LoRA$_3$'' or ``SFT$_3$'').
    \item If all three templates $p_1$, $p_2$, and $p_3$ are used, we indicate prompt augmentation by using ``LoRA$_{123}$'' or ``SFT$_{123}$''.
\end{itemize}

\subsubsection{Hyperparameters}
For LoRA fine-tuning, the default learning rate is set to $1\mathrm{e}{-4}$ and the batch size to 16. We then perform hyperparameter tuning on both of them while fixing the LoRA rank $r$ to 16. Training is conducted for up to 10 epochs, within which all models converge. All experiments are conducted using two random seeds: 42 and 1902582. Random seed 1902582 corresponds to the default setting used in ArcheType’s official implementation. We additionally include random seed 42 and report the average results across the two seeds to control for randomness-induced variance in the results.

\subsubsection{Evaluation Metrics}
While our task is modeled as a text generation problem where the model is prompted with an instruction and produces a label as free-form output, we adopt standard classification-based metrics to evaluate performance. Specifically, we report \textbf{weighted} F1 scores, computed over the predicted labels and their ground truth references. Recall from Section~\ref{sec:preliminaries} that $\mathcal{L}$ denotes the set of all possible semantic labels. We formally define the per-label F1 and weighted F1 scores as follows:

\begin{itemize}
  \item \textbf{Per-label F1 Score} is calculated based on the counts of true positives (TP), false positives (FP), and false negatives (FN) with respect to a specific semantic label. The formula to calculate the per-label F1 score is:
  \[
  \text{F1}_{\ell} = \frac{2 \cdot \text{TP}_{\ell}}{2 \cdot \text{TP}_{\ell} + \text{FP}_{\ell} +  \text{FN}_{\ell}}, \text{ where } \ell\in\mathcal{L}
  \]

  \item \textbf{Weighted F1} averages the per-label F1 scores, with each label weighted by its frequency in the dataset. This metric balances label-level performance while accounting for the granularity at the column-instance level. Denote $n_\ell$ as the count of label $\ell$ in the dataset, the formula to calculate the weighted F1 score is:
  \[
  \text{Weighted F1} = \sum_{\ell \in \mathcal{L}} \frac{n_\ell}{N} \cdot \text{F1}_\ell, \text{ where } N = \sum_{\ell \in \mathcal{L}} n_\ell
  \]
\end{itemize}

It is important to note that the F1 score measures model performance in label prediction, but does not directly capture the model’s sensitivity to prompt variation, which is our primary focus in this work. Therefore, F1 scores are used as an indirect lens: we evaluate the performance across variant prompt templates and analyze the variation in F1 scores to quantify prompt robustness. A model with consistent F1 scores across templates is considered more stable and less sensitive to prompt phrasing.

\subsubsection{Hardware Configuration}
All experiments are conducted on servers managed by SLURM~\cite{10.1007/10968987_3} with NVIDIA H100 SXM5 GPU (80GB) and 32GB of RAM.

\renewcommand{\arraystretch}{1.10}
\begin{table*}[!t]
\centering
\caption{Weighted F1 scores on SOTAB (left) and VizNet (right) test sets across prompt variants ($p_3$, $p_4$, $p_5$).}
\label{table:sensitivity_small}
\begin{tabular}{lccc|ccc}
\toprule
 & \multicolumn{3}{c|}{\textbf{SOTAB}} & \multicolumn{3}{c}{\textbf{VizNet}} \\
\textbf{Model} & $p_3$ & $p_4$ & $p_5$ & $p_3$ & $p_4$ & $p_5$ \\
\midrule
\midrule
FLAN-T5 Base-FROZEN &  {0.301} &  {0.176} & 0.233 &  0.193 &  0.004 & 0.100 \\
\hline
FLAN-T5 Base-0.005-LoRA$_3$ &  0.536 &  0.402 & 0.482 &  0.470 &  0.200 & 0.410 \\
FLAN-T5 Base-0.01-LoRA$_3$ &  0.579 &  0.443 & 0.511 &  0.533 &  0.317 & 0.494 \\
FLAN-T5 Base-0.02-LoRA$_3$ &  0.615 &  0.509 & 0.600 &  0.602 &  0.359 & 0.547 \\
\hline
\hline
FLAN-T5-XXL-FROZEN &  0.472 &  0.384 & 0.433 &  0.381 &  0.199 & 0.318 \\
\hline
FLAN-T5 XXL-0.005-LoRA$_3$ &  0.674 &  0.596 & 0.640 &  0.595 &  0.522 & 0.553 \\
FLAN-T5 XXL-0.01-LoRA$_3$ &  0.708 &  0.644 & 0.665 &  0.658 &  0.591 & 0.611 \\
FLAN-T5 XXL-0.02-LoRA$_3$ &  0.730 &  0.676 & 0.690 &  0.712 &  0.591 & 0.657 \\
\hline
\hline
FLAN-UL2-FROZEN &  0.493 &  0.411 & 0.444 &  0.361 &  0.253 & 0.322 \\
\hline
FLAN-UL2-0.005-LoRA$_3$ &  0.683 &  0.627 & 0.650 &  0.615 &  0.533 & 0.577 \\
FLAN-UL2-0.01-LoRA$_3$ &  0.722 &  0.673 & 0.693 &  0.686 &  0.611 & 0.633 \\
FLAN-UL2-0.02-LoRA$_3$ &  0.744 &  0.704 & 0.708 &  0.735 &  0.649 & 0.698 \\
\bottomrule
\end{tabular}
\end{table*}

\subsection{Prompt Variation Impact}
\label{subsec:cta-sensitivity}
We begin our analysis by showing that prompt sensitivity occurs in both parameter-frozen models and models fine-tuned with a single prompt template.
The sensitivity issue reveals a fundamental limitation of existing prompt-based CTA approaches, which often overlook the instability introduced by surface-level variations, thereby highlighting the need for methods like ours that explicitly address this problem.

Since this experiment includes parameter-frozen models, we focus on the limited-training-data case for fine-tuning, where models are trained on a small set of prompts formulated using $p_3$ and progressively increasing fractions of labeled columns—0.5\%, 1.0\%, and 2.0\% of the original training dataset. We use stratified sampling when selecting labeled columns to construct training prompts in order to maintain the label distribution in the sampled dataset consistent with that of the original dataset. To evaluate the degree of prompt sensitivity, we report model performance in terms of weighted F1 scores on three test sets constructed using different prompt templates: $p_3$ and two additional variants, $p_4$ from~\cite{10.14778/3665844.3665857} and $p_5$, both of which are distinct from $p_1$ and $p_2$.

\begin{enumerate}
    \item Template $p_4$: \texttt{Pick the column's class. I mean if you want to. It would be cool, I think. Anyway, give it a try, I guess? Here's the column itself! [column values] And, um, here are some column names you could pick from ... <labels> OK, go ahead!}
    \item Template $p_5$: \texttt{Given the input column: [column values], choose the most appropriate label from the following options and return only the label. OPTIONS: <labels>}
\end{enumerate}

We choose these two variants to evaluate the model’s generalization to different prompt designs. Template $p_4$ deliberately introduces distractive and informal language to obscure the task instruction, testing the model’s robustness to noise. Template $p_5$ is a clean paraphrase of $p_1$ and $p_2$, designed to probe sensitivity to subtle rewordings.

It is important to note that enumerating all possible prompt formulations is infeasible, given the unbounded space of natural language. Instead, we aim to represent distinct but practically relevant failure modes: informal phrasing and subtle semantic drift. Together, they span a spectrum of instruction variation observed in real-world prompt usage. By demonstrating consistent performance improvements across both cases, our results offer evidence that the model’s reduced sensitivity generalizes beyond our experiment setup.

We use $p_3$ to construct prompts for the single-prompt-template approach when fine-tuning the models, as it achieved the best performance on FLAN models~\cite{wei2022finetuned} among templates $p_1$ through $p_3$ based on our manual exploration. For evaluation, we construct a test set materialized from $p_3$ to serve as a reference point for assessing prompt sensitivity, in comparison with the test sets generated using $p_4$ and $p_5$.
We then compare parameter-frozen FLAN models with their LoRA fine-tuned counterparts on the test sets constructed from the VizNet and SOTAB benchmarks to demonstrate that prompt sensitivity persists even when the models are adapted to the task domain through fine-tuning.
We report the results in Table~\ref{table:sensitivity_small}.

We observe from the table that prompt sensitivity presents a significant challenge across both CTA benchmarks. In the parameter-frozen setting for direct inference, all models experience substantial drops in weighted F1 scores when evaluated on prompt variants $p_4$ and $p_5$ compared to $p_3$, suggesting that pre-trained LLMs may exhibit significantly different responses even when the prompts are semantically similar. Among the three models, FLAN-T5 Base exhibits the most pronounced fluctuations, likely due to the smallest parameter scale. However, even larger models remain susceptible. For example, despite its 11B parameters which are much more than the 0.25B in FLAN-T5 Base, the parameter-frozen FLAN-T5 XXL achieves a weighted F1 of 0.381 on $p_3$ in VizNet, but its performance drops to 0.199 on $p_4$ and 0.318 on $p_5$ when evaluated under direct inference with these prompt templates.

LoRA-based fine-tuning substantially improves model performance on test instances constructed from prompt template $p_3$, as the same prompt pattern is used to generate training examples for fine-tuning the models. As in Table~\ref{table:sensitivity_small}, fine-tuning over just 0.5\% of the training set is sufficient to boost the weighted F1 scores of all FLAN models by at least 20 percentage points on both benchmarks.
However, the performance gap remains evident when switching to the test sets built from the other two prompt templates: most fine-tuned models fail to achieve an F1 score on $p_4$ within four percentage points of their respective $p_3$ scores, indicating that fine-tuned models remain highly sensitive to variations and noise in the prompts. This performance degradation trend persists, albeit to a lesser extent, when the more formally worded prompt $p_5$ is used during inference. These results indicate that although single-prompt-template fine-tuning improves model performance on the CTA task, it does not mitigate the prompt sensitivity issue.

\subsection{Augmentation Improvement}
\label{subsec:aug_improvement}
In this section, we analyze the impact of the augmentation strategy introduced in Section~\ref{subsec:augmentation}. In particular, we aim to investigate whether augmenting the training data with diverse prompt templates improves model robustness to prompt variation. 
We first compare inference results from models fine-tuned on data generated with $p_3$ against those fine-tuned on data augmented with $p_1$, $p_2$, and $p_3$, using 0.5\%, 1\%, and 2\% of the training sets. The weighted F1 scores are reported in Table~\ref{table:augmentation_small}.

\begin{table*}[!t]
\centering
\caption{Weighted F1 scores of models fine-tuned with $p_3$ only versus those fine-tuned with prompt augmentation.} 
\label{table:augmentation_small}
\begin{tabular}{lccc|ccc}
\toprule
 & \multicolumn{3}{c|}{\textbf{SOTAB}} & \multicolumn{3}{c}{\textbf{VizNet}} \\
\textbf{Model} & $p_3$ & $p_4$ & $p_5$ & $p_3$ & $p_4$ & $p_5$ \\
\midrule
\midrule
FLAN-T5 Base-0.005-LoRA$_3$ & 0.536 & 0.402 & 0.482 & 0.470 & 0.200 & 0.410 \\
FLAN-T5 Base-0.005-LoRA$_{123}$ & 0.596 & 0.563 & 0.589 & 0.519 & 0.512 & 0.517 \\
\hline
FLAN-T5 Base-0.01-LoRA$_3$ & 0.579 & 0.443 & 0.511 & 0.533 & 0.317 & 0.494 \\
FLAN-T5 Base-0.01-LoRA$_{123}$ & 0.623 & 0.611 & 0.623 & 0.601 & 0.595 & 0.597 \\
\hline
FLAN-T5 Base-0.02-LoRA$_3$ & 0.615 & 0.509 & 0.600 & 0.602 & 0.359 & 0.547 \\
FLAN-T5 Base-0.02-LoRA$_{123}$ & 0.661 & 0.637 & 0.660 & 0.646 & 0.638 & 0.645 \\
\hline
\hline
FLAN-T5 XXL-0.005-LoRA$_3$ & 0.674 & 0.596 & 0.640 & 0.595 & 0.522 & 0.553 \\
FLAN-T5 XXL-0.005-LoRA$_{123}$ & 0.701 & 0.697 & 0.695 & 0.630 & 0.629 & 0.624 \\
\hline
FLAN-T5 XXL-0.01-LoRA$_3$ & 0.708 & 0.644 & 0.665 & 0.658 & 0.591 & 0.611 \\
FLAN-T5 XXL-0.01-LoRA$_{123}$ & 0.735 & 0.728 & 0.727 & 0.704 & 0.698 & 0.701 \\
\hline
FLAN-T5 XXL-0.02-LoRA$_3$ & 0.730 & 0.676 & 0.690 & 0.712 & 0.591 & 0.657 \\
FLAN-T5 XXL-0.02-LoRA$_{123}$ & 0.751 & 0.734 & 0.741 & 0.747 & 0.737 & 0.748 \\
\hline
\hline
FLAN-UL2-0.005-LoRA$_3$ & 0.683 & 0.627 & 0.650 & 0.615 & 0.533 & 0.577 \\
FLAN-UL2-0.005-LoRA$_{123}$ & 0.708 & 0.697 & 0.699 & 0.664 & 0.654 & 0.661 \\
\hline
FLAN-UL2-0.01-LoRA$_3$ & 0.722 & 0.673 & 0.693 & 0.686 & 0.611 & 0.633 \\
FLAN-UL2-0.01-LoRA$_{123}$ & 0.735 & 0.729 & 0.726 & 0.706 & 0.701 & 0.704 \\
\hline
FLAN-UL2-0.02-LoRA$_3$ & 0.744 & 0.704 & 0.708 & 0.735 & 0.649 & 0.698 \\
FLAN-UL2-0.02-LoRA$_{123}$ & 0.752 & 0.735 & 0.742 & 0.785 & 0.774 & 0.780 \\
\bottomrule
\end{tabular}
\end{table*}

The table indicates that our augmentation substantially reduces performance fluctuations caused by using different test prompts for the CTA task. Specifically, no model shows more than a 3.3-percentage-point variation in weighted F1 scores across all templates on either dataset. In addition, we observe the same trend regardless of which test prompt we use: fine-tuning with multi-prompt ($p_1$+$p_2$+$p_3$) augmentation improves weighted F1 score by at least 1.2 percentage points compared with fine-tuning on a single prompt pattern ($p_3$ only). These two observations suggest that our strategy contributes to stabilizing models even in the presence of noise. 

However, it remains unclear whether the observed improvement in mitigating prompt sensitivity stems from the absolute scale of labeled data or from training data augmentation using prompt variants. To further examine this, we move to the abundant-training-data case and design the following comparison: models fine-tuned using the entire SOTAB training set with only $p_3$, versus models trained on one third of the training set but augmented with $p_1$ through $p_3$. The total number of training prompts is therefore controlled to be identical across the two experimental settings. This setup allows us to examine whether the model continues to exhibit prompt sensitivity at inference time after being exposed to a larger number of labeled column examples but within a single prompt template and whether the proposed augmentation method can mitigate prompt sensitivity while requiring fewer labeled columns by constructing augmented prompts with multiple variant templates.

\begin{figure}[!b]
    \centering
    \includegraphics[width=0.47\textwidth]{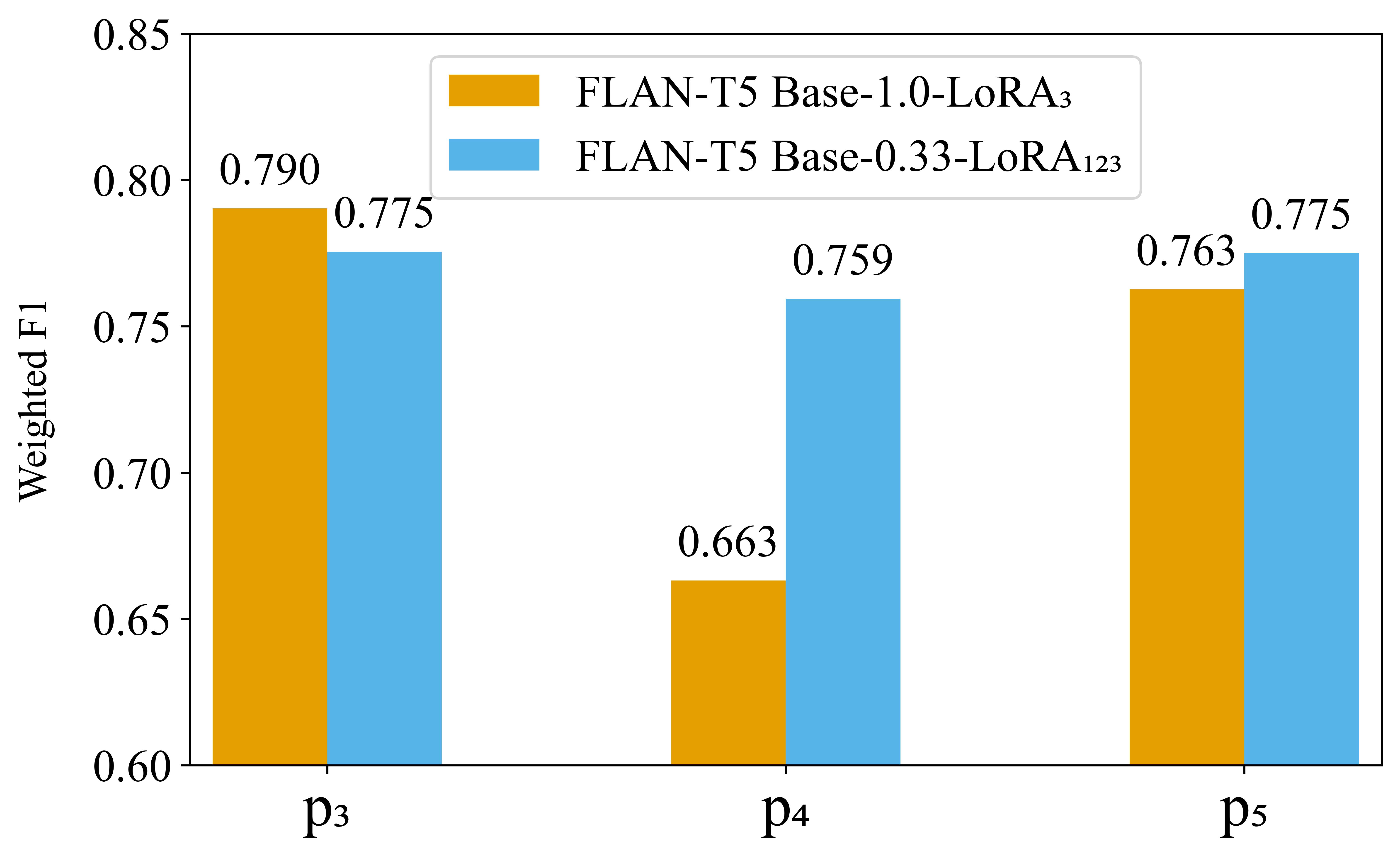}
    \caption{
        Weighted F1 scores of FLAN-T5 Base LoRA fine-tuned with and without prompt augmentation on the SOTAB dataset.
    }
    \label{fig:f1-t5base}
\end{figure}

\begin{figure}[!htbp]
    \centering
    \includegraphics[width=0.47\textwidth]{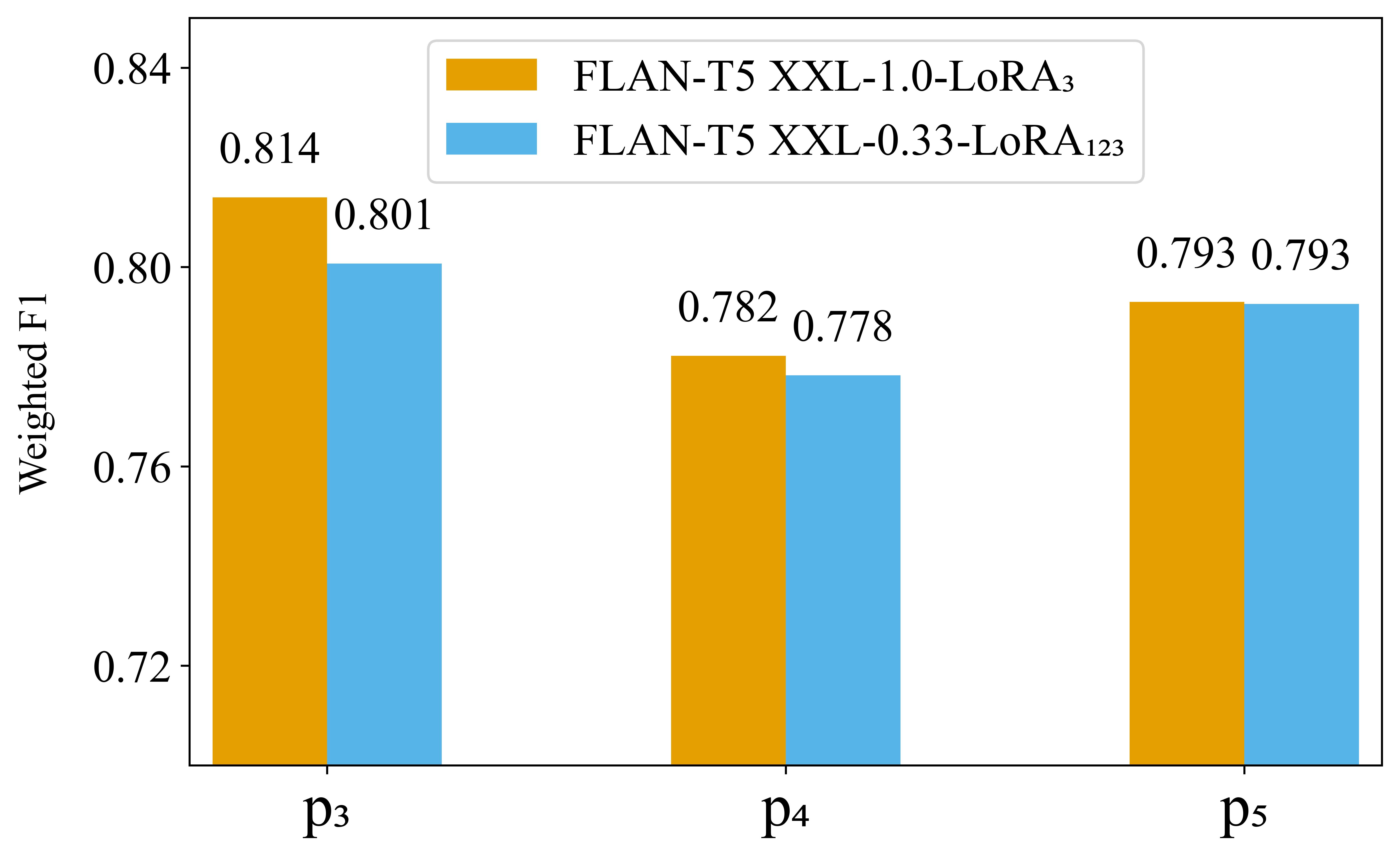}
    \caption{
        Weighted F1 scores of FLAN-T5 XXL LoRA fine-tuned with and without prompt augmentation on the SOTAB dataset.
    }
    \label{fig:f1-t5xxl}
\end{figure}

\begin{figure}[!htbp]
    \centering
    \includegraphics[width=0.47\textwidth]{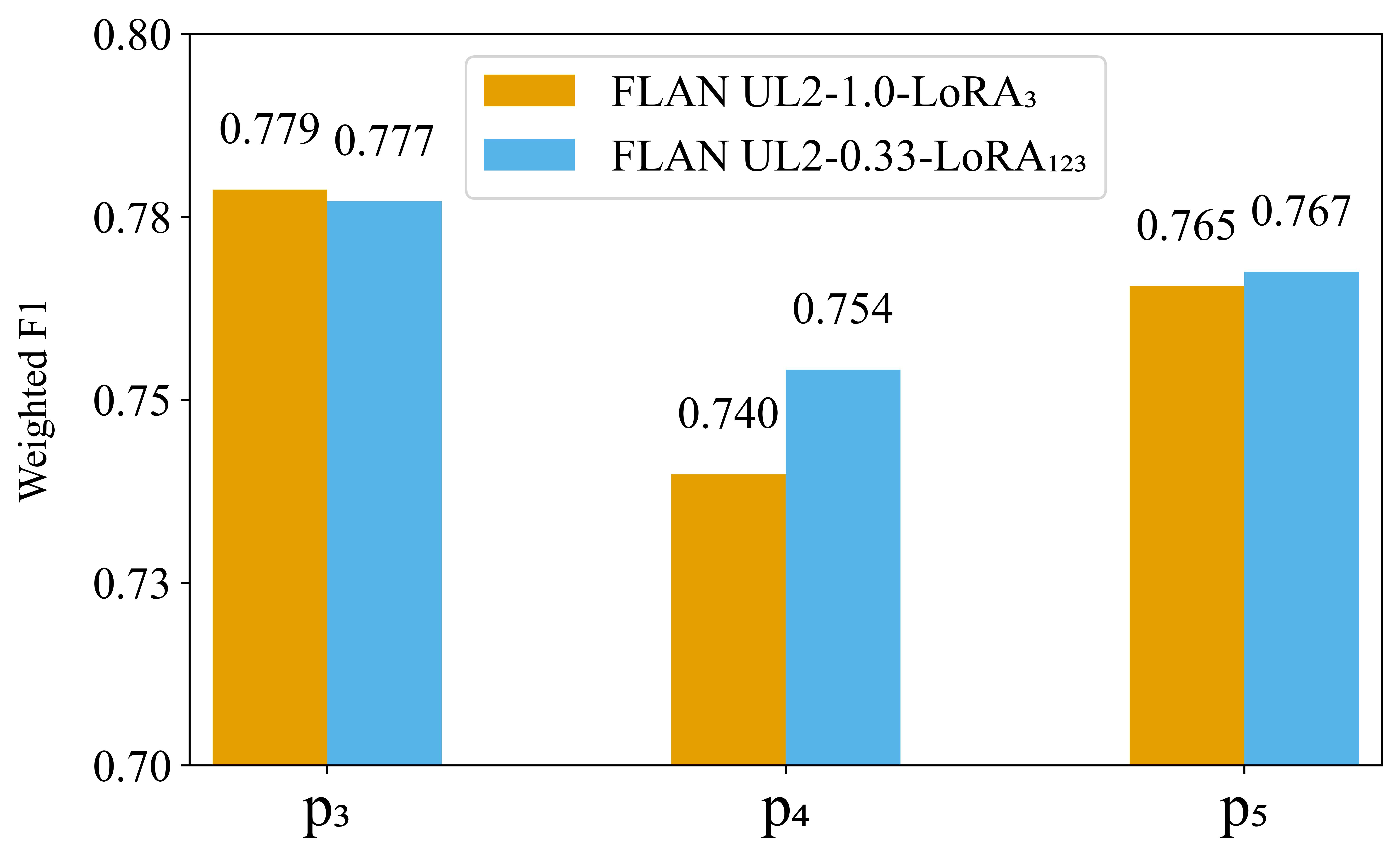}
    \caption{
        Weighted F1 scores of FLAN-UL2 LoRA fine-tuned with and without prompt augmentation on the SOTAB dataset.
    }
    \label{fig:f1-ul2}
\end{figure}

In Figures 5–7, we observe that models fine-tuned using only $p_3$ achieve the best performance on the test set constructed with $p_3$, as it is the same prompt template used to construct both the fine-tuning and test sets. However, when we examine the other test sets built with $p_4$ and $p_5$, the models fine-tuned solely on $p_3$ exhibit greater sensitivity to variations in test prompts compared to models fine-tuned on training prompts constructed from one-third of the column examples but augmented with the other two prompt templates, $p_1$ and $p_2$. In other words, the performance difference on test sets $p_4$ and $p_5$ relative to test set $p_3$ is smaller for the models fine-tuned with prompt augmentation.
For instance, even for FLAN-UL2, which has the largest number of parameters among the three FLAN models compared here, the weighted F1 score can still vary by nearly four percentage points when the model is fine-tuned using only $p_3$. In contrast, prompt augmentation reduces this difference to 2.3 percentage points, despite using only one-third of the labeled column examples to build the training prompts. This demonstrates the effectiveness of our augmentation method in alleviating the prompt sensitivity issue. Furthermore, the results suggest that models trained on just one-third of the labeled column examples, when augmented with variant prompt templates, can achieve performance comparable to those trained on the entire dataset, despite using two-thirds fewer labeled columns. This finding highlights that our augmentation strategy can effectively compensate for limited labeled data in fine-tuning.

\subsection{Case Study: Full-Model Fine-tuning}
\label{subsec:case_llama}
\begin{figure}[!htbp]
    \centering
    \includegraphics[width=0.47\textwidth]{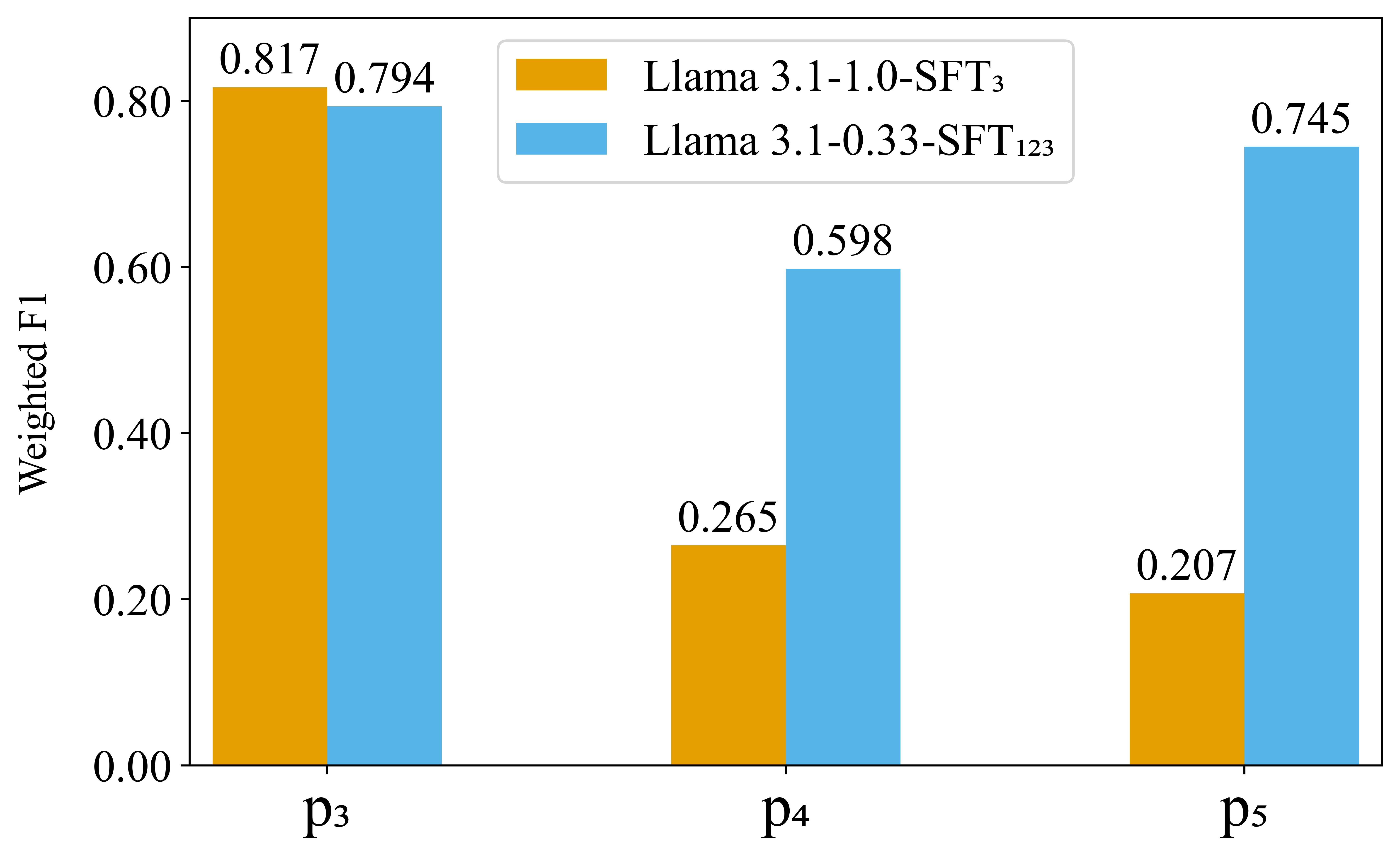}
    \caption{
        Weighted F1 scores of Llama 3.1 Instruct fine-tuned with and without prompt augmentation on the SOTAB dataset
    }
    \label{fig:f1-llama}
\end{figure}

We have previously shown that generative models exhibit unstable performance across different templates in CTA tasks, even when fine-tuned without augmentation. However, it remains unclear whether this instability may also be attributed to the inherent limitations of the FLAN models~\cite{wei2022finetuned} or the restricted number of tunable parameters imposed by LoRA~\cite{hu2022lora}. To further investigate this question, we fine-tune Llama 3.1 8B Instruct~\cite{DBLP:journals/corr/abs-2407-21783} by updating all of its parameters, both with and without prompt augmentation. Again, to keep the total number of fine-tuning examples consistent, we use one-third of the SOTAB training data for prompt augmentation and the full training set for standard fine-tuning. We then evaluate the models on the SOTAB test set and report their weighted F1 scores in Figure~\ref{fig:f1-llama}.

The figure shows that the model exhibits even greater sensitivity to prompt variation compared to those fine-tuned with LoRA when both are trained solely on $p_3$. The average F1 drop on $p_4$ and $p_5$ relative to $p_3$ reaches 58.06 percentage points. The model surprisingly performs slightly worse on $p_5$ than on $p_4$, which is an unexpected outcome, as $p_4$ was deliberately perturbed with irrelevant words, whereas $p_5$ retains a clear instructional structure. Nevertheless, our augmentation remains effective: it helps the model restore the F1 score on $p_5$ to a level comparable to that of $p_3$ and improves performance on $p_4$ by 33 percentage points, when comparing to the model fine-tuned with $p_3$ only. These findings highlight the continued effectiveness of our prompt augmentation strategy. Despite being fully fine-tuned with all parameters updated, the more recent Llama model still exhibits vulnerability to prompt variation, suggesting that such sensitivity persists with up-to-date pre-training techniques and sufficient amount of trainable parameters. The fact that our augmentation approach continues to stabilize performance in this setting highlights its potential as a broadly applicable technique for improving robustness to prompt variation.

\subsection{Discussion and Takeaways}

Our experimental findings offer several key insights into the behavior and robustness of generative large language models with respect to CTA tasks, especially under prompt variations.

\paragraph{Prompt Sensitivity Is a Persistent Challenge}  
Across both parameter-frozen and fine-tuned settings, we observe consistent drops in weighted F1 scores when models are exposed to variant prompt templates. This instability remains significant even for instruction-tuned models such as the FLAN and Llama models. These results indicate that prompt phrasing is not merely a surface-level variation, but a critical factor that can substantially influence label prediction performance.

\paragraph{Fine-Tuning Does Not Resolve Fragility}  
Although LoRA-based fine-tuning improves weighted F1 scores on $p_3$, it does not eliminate degradation when evaluated on $p_4$ and $p_5$ which are not used for training. In many cases, the drop in performance exceeds 4 percentage points, suggesting that the model, even after fine-tuning, is susceptible to prompt phrasing.

\paragraph{Prompt Augmentation Enhances Generalization and Data Efficiency}  
Our prompt augmentation strategy significantly reduces the performance gap across test instances constructed from different prompt templates. Even when fine-tuned on a prompt-augmented training set built from only 0.5–2\% of the original data, models exhibit improved stability across prompt designs and produce more accurate results, as evidenced by gains in weighted F1 scores. Moreover, fine-tuning on only one-third of the original training data augmented with three variant prompt templates achieves performance comparable to full-data single-prompt-template fine-tuning. This finding underscores the data efficiency benefit introduced by prompt variation, which is particularly valuable when labeled columns are difficult to obtain.

\paragraph{Prompt Augmentation Is a Model-Agnostic, Scalable Solution}  
We evaluate models of varying sizes and fine-tuning scopes and find that prompt sensitivity persists across all configurations where the model is fine-tuned with respect to a single prompt pattern. These results indicate that neither model scale nor the extent of parameter updates is sufficient to ensure robustness. In contrast, our prompt augmentation strategy consistently improves model performance and reduces response variability across test sets constructed from variant prompts. In conjunction with LoRA-based fine-tuning, it provides a lightweight, scalable, and model-agnostic solution for improving model robustness in prompt-based classification tasks.
\section{Conclusion and Future Work}
\label{sec:conclusion}

This paper addresses the challenge of prompt sensitivity in column type annotation tasks. We propose a systematic prompt augmentation strategy and employ a lightweight, parameter-efficient fine-tuning framework based on Low-Rank Adaptation (LoRA)~\cite{hu2022lora}. Empirical results on two real-world datasets, VizNet~\cite{10.1145/3290605.3300892} and SOTAB~\cite{Korini2022SOTABTW}, demonstrate that the proposed approach significantly enhances model robustness. Specifically, we find that prompt augmentation not only stabilizes responses across varied instruction templates at inference time but also enables the model to produce overall more accurate annotation results compared to single-prompt-template fine-tuning.

Our study also reveals that prompt sensitivity persists across all experimental settings involving single-template fine-tuning, regardless of whether full model parameters are updated or only the LoRA adapters, and independent of the underlying model choice. This suggests that the prompt sensitivity issue is not a consequence of limited fine-tuning capacity but rather a more fundamental challenge, highlighting the broader need for training strategies that generalize across input perturbations.

A promising direction for future research is to extend our augmentation framework to incorporate unlabeled columns into the fine-tuning data. A natural extension is to incorporate consistency regularization~\cite{tarvainen2017mean, laine2017temporal, 10.1145/3626766} over unlabeled columns, by encouraging models to produce consistent predictions across multiple prompt variants of the same input. This could be achieved by minimizing the divergence between model outputs over different augmented prompts, without relying on ground truth labels. Such an approach would align well with semi-supervised or unsupervised learning paradigms and would further enhance the practicality of CTA systems in low-resource or rapidly evolving data environments.


\bibliographystyle{ACM-Reference-Format}
\bibliography{sample}

\end{document}